\documentclass[11pt]{article}

\usepackage{amsmath}
\usepackage{amssymb}
\usepackage{hyperref}

\usepackage{xcolor}
\usepackage{bbm}
\usepackage{maybemath}
\newcommand{\plus}{{\mbox{{\bf{\small +}}}}}

\definecolor{garrosgreen}{rgb}{0.1, 0.4, 0.1}
\definecolor{dartmouthgreen}{rgb}{0.05, 0.5, 0.06}
\definecolor{duelferred}{rgb}{0.7, 0.2, 0.1}
\definecolor{cambridgeblue}{rgb}{0.1, 0.3, 1.0}
\definecolor{oxfordblue}{rgb}{0.05, 0.2, 0.7}

\newcommand{\ee}{{\mathrm e}}
\newcommand{\ii}{{\mathrm i}}
\newcommand{\dd}{{\mathrm d}}

\newcommand{\calC}{\mathcal{C}}
\newcommand{\calD}{\mathcal{D}}
\newcommand{\calL}{\mathcal{L}}

\newcommand{\rmT}{{\rm T}}

\begin{document}

\title{Antimatter Gravity: Second Quantization and Lagrangian Formalism}

\author{Ulrich D. Jentschura\\
\small
Department of Physics, Missouri University\\
\small
of Science and Technology, Rolla, Missouri 65409, USA; ulj@mst.edu\\
\small
MTA--DE Particle Physics Research Group,\\
\small
P.O.~Box 51, H--4001 Debrecen, Hungary\\
\small
MTA Atomki, P.O.~Box 51, H--4001 Debrecen, Hungary}

\maketitle

\begin{abstract}
The application of the CPT theorem to an apple
falling on Earth leads to the description
of an anti-apple falling on anti--Earth
(not on Earth). On the microscopic level,
the Dirac equation in curved space-time
simultaneously describes spin-$1/2$ particles 
and their antiparticles coupled to the same curved space-time
metric (e.g., the metric describing the 
gravitational field of the Earth).
On the macroscopic level, the electromagnetically 
and gravitationally coupled Dirac equation
therefore describes apples and anti-apples,
falling on Earth, simultaneously.
A particle-to-antiparticle transformation of the 
gravitationally coupled 
Dirac equation therefore yields information on the 
behavior of ``anti-apples on Earth''.
However, the problem is exacerbated by the fact that the 
operation of charge conjugation is much more complicated
in curved as opposed to flat space-time.
Our treatment is based on second-quantized field
operators and uses the Lagrangian formalism.
As an additional helpful result, prerequisite to our calculations,
we establish the general 
form of the Dirac adjoint in curved space-time. 
On the basis of a theorem,
we refute the existence of tiny, but potentially
important, particle-antiparticle symmetry breaking
terms whose possible existence has been investigated 
in the literature. Consequences for antimatter 
gravity experiments are discussed.
\end{abstract}

\noindent Keywords: General Relativity, Antimatter Gravity,
Antiparticles, CPT Symmetry, Spin Connection

%
%
\section{Introduction}
\label{sec1}

It is common wisdom in atomic physics that 
the Dirac equation describes particles and
antiparticles simultaneously,
and that the negative-energy solutions
of the Dirac equation have to be 
reinterpreted in terms of particles that 
carry the opposite charge as compared to 
particles, and whose numerical value
of the energy $E$ is equal to the negative
value of the physically observed energy~\cite{ItZu1980}.
Based on the Dirac equation, the existence of the 
positron was predicted, followed by 
its experimental detection in 1933, by Anderson~\cite{An1933}.
If we did not reinterpret the negative-energy
solutions of the Dirac equation,
then the helium atom would be unstable against
decay into a state where the two
electrons perform quantum jumps into 
continuum states~\cite{BrRa1951}.
One of the electrons would jump into the positive-energy
continuum, the other, into the negative-energy 
continuum, with the sum of the energies of the two continuum
states being equal to the sum of the two 
bound-state energies of the helium atom
from which the transition started~\cite{BrRa1951,JaBuLK1997,Ma2016}.

The absolute necessity to reinterpret the 
negative-energy solutions of the Dirac equation
as antiparticle wave functions,
i.e., the necessity to interpret positive-energy and 
negative-energy solutions of one single equation as
describing two distinct particles,
hints at the possibility to use the 
Dirac equation as a bridge to the 
description of the gravitational 
interaction of antimatter.
Namely, if the Dirac equation is being
coupled to a gravitational field, then, since it
describes particles and antiparticles simultaneously,
the Dirac equation offers us an additional dividend:
In addition to describing the gravitational interaction
of particles, the Dirac equation automatically
couples the antiparticle (the ``anti-apple''), which is described
by the same equation, to the gravitational field, too.

Corresponding investigations have been initiated in a series
of recent publications~\cite{JeNo2013pra,Je2013,Je2018geonium,Je2019ijmpa}.
One may ask whether the dynamics of particles and antiparticles
differ in a central, static, gravitational
field, in first approximation, but also, if there
are any small higher-order effects breaking
the particle-antiparticle symmetry
under the gravitational interaction.
The first of these questions has been answered
in Refs.~\cite{JeNo2013pra,Je2013,Je2018geonium}, 
with the result being that the 
Dirac particle and antiparticle behave exactly the 
same in a central gravitational field,
due to a perfect particle-antiparticle symmetry
which extends to the relativistic and
curved-space-time corrections to the equations of motion.

In order to address the second question, it is 
necessary to perform the full particle-to-antiparticle
symmetry transformation of the Dirac formalism,
in an arbitrary (possibly dynamic)
curved-space-time-background.
This transformation is most stringently carried out on the 
level of the Lagrangian formalism.
A preliminary result has recently been 
published in Ref.~\cite{Je2019ijmpa}, where a relationship
was established between the positive-energy
and negative-energy solutions of the Dirac equation
in an arbitrary dynamics curved-space-time-background.
However, the 
derivation in Ref.~\cite{Je2019ijmpa} is based
on a first-quantized formalism, which lacks 
the unified description in terms of the field operator.
The field operator
comprises {\em all} (as opposed to {\em any}) 
solution of the gravitationally (and electromagnetically) 
coupled Dirac equation. 
In general, a satisfactory description of antiparticles,
in the field-theoretical context, necessitates 
a description in terms of particle- and antiparticle 
creation and annihilation processes, and therefore,
the introduction of a field operator.
In consequence, the investigation~\cite{Je2019ijmpa} 
is augmented here on the 
basis of a transformation of the entire
Lagrangian density, which can be expressed
in terms of the charge-conjugated (antiparticle)
bispinor wave function, and generalized to the level
of second quantization.
The origin~\cite{Je2019ijmpa} 
of a rather disturbing minus sign which
otherwise appears in the Lagrangian formalism
upon charge conjugation in first quantization
will be addressed. The use of the Lagrangian formalism 
necessitates a definition of the Dirac adjoint
in curved space-times. As a spin-off result of the 
augmented investigations reported here, 
we find the general form of the Dirac adjoint
in curved space-times, in the Dirac representation
of $\gamma$ matrices.

According to Ref.~\cite{HoEtAl1991},
the role of the CPT transformation in gravity needs
to be considered with care: {\em A priori},
a CPT transformation of a physical system
consisting of an apple falling on Earth would 
describe the fall of an anti-apple on anti-Earth.
Key to our considerations is the fact that,
on the microscopic level, the Dirac equation 
applies (for one and the same space-time metric)
to both particles and antiparticles
simultaneously (this translates, on the macroscopic level,
to ``apples'' as well as ``anti-apples'').
This paper is organized as follows:
We investigate the general form of 
the Dirac adjoint in Sec.~\ref{sec2},
present our theorem in Sec.~\ref{sec3},
and in Sec.~\ref{sec4}, we provide an overview of 
connections to new forces and CPT violating parameters,
Conclusions are reserved for Sec.~\ref{sec5}.

%
%
\section{Dirac Adjoint for Curved Space--Times}
\label{sec2}

In order to properly write down the Lagrangian of a 
Dirac particle in a gravitational field,
we first need to generalize the concept of the Dirac adjoint to curved
space-times. We recall that the 
Dirac adjoint transforms with the inverse of 
the Lorentz transform as compared to the original Dirac spinor.
A general spinor Lorentz transformation $S(\Lambda)$ is given as follows,
\begin{equation}
\label{defSlorentz}
S(\Lambda) = \exp\left(-\frac{\ii}{4} \, 
\epsilon^{AB} \, 
\sigma_{AB} \right) \,,
\qquad
\sigma_{AB} = \frac{\ii}{2} \, 
\left[ \gamma^A, \gamma^B \right] \,,
\qquad
A, B = 0,1,2,3 \,.
\end{equation}
Note that the generator parameters $\epsilon^{AB}=-\epsilon^{BA}$,
for local Lorentz transformations, 
can be coordinate-dependent.
In the following, capital Roman letters $A,B,C,\dots = 0,1,2,3$ 
refer to Lorentz indices in a local freely falling coordinate system.
The (flat-space) Dirac matrices $\gamma^A$ are 
assumed to be taken in the Dirac representation~\cite{ItZu1980},
\begin{equation}
\gamma^0 =
\left( \begin{array}{cc}
\mathbbm{1}_{2 \times 2}  & 0 \\
0 & \mathbbm{1}_{2 \times 2}
\end{array} \right) \,,
\qquad
\vec\gamma =
\left( \begin{array}{cc}
0 & \vec\sigma \\
-\vec\sigma & 0 \\
\end{array} \right) \,.
\end{equation}
Here, the vector of Pauli spin matrices is denoted as $\vec\sigma$.
In consequence, the spin matrices $\sigma_{AB}$
are the flat-space spin matrices.
The spin matrices fulfill the commutation relations
\begin{equation}
\label{commutator_sigma}
[ \tfrac12 \sigma^{C D}, \tfrac12 \sigma^{E F}] =
\ii \left( g^{C F} \tfrac12 \sigma^{D E}
+ g^{D E} \tfrac12 \sigma^{C F}
- g^{C E} \tfrac12 \sigma^{D F}
- g^{D  F} \tfrac12 \sigma^{CE} \right) \,.
\end{equation}
These commutation relations, we should note in passing,
are completely analogous to those fulfilled by the 
matrices $\mathbbm{M}_{A B}$ that generate 
(four-)vector local Lorentz transformations.
As is well known, the latter have the components (denoted by indices $C$ and $D$)
\begin{equation}
\label{Malphabeta}
{(\mathbbm{M}_{A B})^C}_D =
{g^C}_A \, g_{DB} -
{g^C}_B \, g_{DA} \,.
\end{equation}
The vector local Lorentz transformation $\Lambda$ 
with components ${\Lambda^C}_D$ is obtained
as the matrix exponential
\begin{equation}
\label{defMlorentz}
{\Lambda^C}_D =
{\left( 
\exp\left[ \frac12 \, \epsilon^{AB} \, 
\mathbbm{M}_{AB}\right] 
\right)^C}_D \,.
\end{equation}
The algebra fulfilled by the $\mathbbm{M}$ matrices is well known to be
\begin{equation}
\label{commutator_m}
[ \mathbbm{M}^{C D}, \mathbbm{M}^{E F}] = 
g^{C F} \; \mathbbm{M}^{D E}
+ g^{D E} \; \mathbbm{M}^{C F}
- g^{C E} \; \mathbbm{M}^{D F}
- g^{D  F} \; \mathbbm{M}^{CE} \,.
\end{equation}
The two algebraic relations~\eqref{commutator_sigma}
and~\eqref{commutator_m} are equivalent if one replaces
\begin{equation}
\mathbbm{M}^{C D} \to
-\frac{\ii}{2} \, \sigma^{CD} \,,
\end{equation}
which leads from Eq.~\eqref{defSlorentz} to Eq.~\eqref{defMlorentz}.
Under a local Lorentz transformation, a Dirac spinor transforms as
\begin{equation}
\label{trafoSpsi}
\psi'(x') = S(\Lambda) \, \psi(x) \,.
\end{equation}
In order to write the Lagrangian, one needs to define the Dirac adjoint 
in curved space-time. In order to address this question,
one has to remember that in flat-space-time, the 
Dirac adjoint $\overline \psi(x)$ 
is defined in such a way that is transforms with the 
inverse of the spinor Lorentz transform as compared to $\psi(x)$,
\begin{equation}
\label{cayetano}
\overline\psi'(x') 
= \overline\psi(x) \, S(\Lambda^{-1})
= \overline\psi(x) \, [ S(\Lambda) ]^{-1} \,.
\end{equation}
The problem of the definition of $\overline \psi(x)$
in curved space-time is sometimes treated in the literature
in a rather cursory fashion~\cite{BrWh1957}.
Let us see if in curved space-time, we can use the {\em ansatz}
\begin{equation}
\label{ansatz}
\overline\psi(x) \, = \psi^\plus(x) \, \gamma^0 \,,
\end{equation}
with the same flat-space $\gamma^0$ as is used in the flat-space
Dirac adjoint.
In this case,
\begin{equation}
\overline\psi'(x') = 
\psi^\plus(x') \, S^\plus(\Lambda) \, \gamma^0 
= \left( \psi^\plus(x') \, \gamma^0 \right) \; 
\left[ \gamma^0 \, S^\plus(\Lambda) \, \gamma^0 \right] \,,
\end{equation}
To first order in the Lorentz generators $\epsilon_{AB}$, we have
indeed,
\begin{equation}
\label{salvation}
\gamma^0 \, S^\plus(\Lambda) \, \gamma^0 
= 1 + \frac{\ii}{4} \, \epsilon^{AB} \, \gamma^0 \, \sigma^\plus_{AB} \, \gamma^0 
= 1 + \frac{\ii}{4} \, \epsilon^{AB} \, \sigma_{AB}  
= [ S(\Lambda) ]^{-1} \,,
\end{equation}
where we have used the identity
\begin{align}
\label{sigmaplusAB}
\sigma^\plus_{AB} 
=& \; -\frac{\ii}{2} \, [ \gamma^\plus_B, \gamma^\plus_A ] 
= -\frac{\ii}{2} \, \gamma^0  \;
[ \gamma^0 \gamma^\plus_B \gamma^0, \;
\gamma^0 \gamma^\plus_A \gamma^0 ] \; \gamma^0 
\nonumber\\[0.1133ex]
=& \; -\frac{\ii}{2} \, \gamma^0 \;
[ \gamma_B, \; \gamma_A ] \; \gamma^0 
= - \gamma^0 \; \sigma_{BA} \; \gamma^0 
= \gamma^0 \; \sigma_{AB} \; \gamma^0 \, .
\end{align}
It is easy to show that Eq.~\eqref{salvation} generalizes
to all orders in the $\epsilon^{AB}$ parameters,
which justifies our {\em ansatz} given in Eq.~\eqref{ansatz}.
The result is that the flat-space $\gamma^0$ matrix can be 
used in curved space, just like in flat space, in order to construct the Dirac
adjoint. The Dirac adjoint spinor transforms with the 
inverse spinor representation of the Lorentz group [see Eq.~\eqref{cayetano}].

%
%
\section{Lagrangian and Charge Conjugation}
\label{sec3}

Equipped with an appropriate form of the Dirac adjoint 
in curved space-time, we start from the Lagrangian 
density~\cite{FoIw1929,Fo1929,Fo1929crasp,BrWh1957,%
Bo1975prd,SoMuGr1977,Iv1969a,Iv1969b,IvMiVl1985,Bo2011}
\begin{equation}
\label{ord}
\calL = \overline \psi(x)
\left[ \overline\gamma^\mu 
\left\{ \ii \left( \partial_\mu - \Gamma_\mu \right) - e \, A_\mu \right\} - 
m_I \right] \, \psi(x) \,,
\end{equation}
Here, the $A_\mu$ field describes the four-vector potential
of the electromagnetic field,
while the $\Gamma_\mu$ matrices describe the spin connection.
\begin{equation}
\label{Gamma}
\Gamma_\mu = \frac{\ii}{4} \, \omega_\mu^{AB} \, \sigma_{AB} \,,
\qquad
\omega^{AB}_\mu = e^A_\nu \nabla_\mu e^{\nu B} \,,
\qquad
\nabla_\mu e^{\nu B} =
\partial_\mu e^{\nu B} +
\Gamma^\nu_{\mu\rho} \, e^{\rho B} \,.
\end{equation}
For the form of the covariant coupling, we refer to 
Eqs.~(3.129) and (3.190) of Ref.~\cite{Bo2011}.
In the above equations, capital Roman 
indices $A,B,C,\dots = 0,1,2,3$
refer to a freely falling coordinate system (a Lorentz index), 
while Greek indices $\mu,\nu,\rho,\dots = 0,1,2,3$ 
refer to an external coordinate system (an Einstein index).

We shall attempt to derive the particle-antiparticle symmetry on
the level of a transformation of the Lagrangian.
In comparison to textbook treatments (see, e.g., 
pp.~89 ff.~and 263 ff.~of Ref.~\cite{AkBe1969},
p.~70 of Ref.~\cite{PeSc1995},
p.~66 of Ref.~\cite{Ga1975},
pp.~89 ff.~and 263 ff.~of Ref.~\cite{Ga1975},
p.~142 of Ref.~\cite{JaRo1980},
p.~218 of Ref.~\cite{LaPa2011},
p.~67 of Ref.~\cite{BjDr1964},
p.~116 of Ref.~\cite{BjDr1965},
p.~320 of Ref.~\cite{BoLoTo1975},
p.~153 of Ref.~\cite{ItZu1980},
and Chap.~7 of Ref.~\cite{Kl2016}),
our derivation is much more involved in view of the 
appearance of the $\Gamma_\mu$ matrices 
which describe the gravitational coupling.
In other words, we note that 
{\em none} of the mentioned standard textbooks of quantum 
field theory discuss the {\em gravitationally}
coupled Dirac equation, and all cited 
descriptions are limited to the flat-space Dirac 
equation, where the role of the 
charge conjugation operation is much easier to 
analyze than in curved space.

The double-covariant coupling to both the 
gravitational as well as the electromagnetic field is
given as follows,
\begin{equation}
\calD_\mu = 
\partial_\mu - \Gamma_\mu + \ii e \, A_\mu =
\nabla_\mu + \ii e \, A_\mu \,,
\end{equation}
where $\nabla_\mu = \partial_\mu - \Gamma_\mu$ is the gravitational
covariant derivative.

As a side remark, we note that gravitational spin connections 
$\Gamma_\mu = \frac{\ii}{4} \, \omega_\mu^{AB} \, \sigma_{AB}$
and other gauge-covariant couplings are unified in the 
so-called spin-charge family theory~\cite{MB2001,MBNi2003,MB2015,MBNi2016,MB2017}
which calls for a unification of all known 
interactions of nature in terms of an $SO(1,13)$ 
overarching symmetry group.
(In the current article, we use the spin connection matrices purely in the 
gravitational context.) The $SO(1,13)$ has a 
25-dimensional Lie group, with 13 boosts and 12 rotations
in the internal space. This provides for enough Lie algebra
elements to describe the Standard Model interactions,
and predict a fourth generation of particles. The spin-charge 
family theory is a significant generalization of 
Kaluza-Klein-type ideas~\cite{Ka1921,Kl1926}.

In the context of the current investigations, though, we 
restrict ourselves to the gravitational spin connection matrices.
In view of the (in general) nonvanishing 
space-time dependence of the Ricci rotation coefficients, 
we can describe the quantum dynamics of 
relativistic spin-$1/2$ particles on the basis of Eqs.~\eqref{ord} and~\eqref{Gamma}.
The $\sigma_{AB}$ matrices defined in Eq.~\eqref{Gamma}
represent the six generators of the spin-$1/2$ representation
of the Lorentz group.

The Lagrangian~\eqref{ord} is Hermitian, and so
\begin{equation}
\label{Lorig}
\calL = \calL^\plus = \psi^\plus(x)
\left[ (\overline\gamma^\mu)^\plus
\left\{ -\ii \overleftarrow \partial_\mu - e \, A_\mu \right\}
- (-\ii) \left(\Gamma_\mu \right)^\plus ( \overline\gamma^\mu)^\plus - 
m_I \right] \, \left[ \, \overline\psi(x) \, \right]^\plus \,.
\end{equation}
An insertion of $\gamma^0$ matrices under use of the identity
$(\gamma^0)^2 = 1$ leads to the relation
\begin{align}
\calL^\plus =& \; \psi^\plus(x) \,  \gamma^0 \,
\left[  \gamma^0 \, (\overline\gamma^\mu)^\plus \,  \gamma^0 
\left\{ -\ii \overleftarrow \partial_\mu - e \, A_\mu \right\}
\right.
\nonumber\\[0.1133ex]
& \; \left. 
+ \ii \left\{ \gamma^0 \, \left(\Gamma_\mu \right)^\plus  \,  \gamma^0 \right\}
\gamma^0 \, (\overline\gamma^\mu)^\plus \,  \gamma^0 - 
m_I \right] \, \gamma^0 \, \left[ \, \overline\psi(x) \, \right]^\plus \,.
\end{align}
Also, we recall that
$\gamma^0 \, (\Gamma_\mu)^\plus \,  \gamma^0 = -\Gamma_\mu$,
because
\begin{equation}
\label{signchange}
\Gamma^\plus_\mu 
= -\frac{\ii}{4} \, \omega_\mu^{AB} \, \sigma^\plus_{AB} 
= -\frac{\ii}{4} \, \omega_\mu^{AB} \, \gamma^0 \, \sigma_{AB} \, \gamma^0 
= - \gamma^0 \, \Gamma_\mu \gamma^0 \,.
\end{equation}
So, the adjoint of the Lagrangian is 
\begin{equation}
\calL^\plus = \psi^\plus(x) \,  \gamma^0 \,
\left[  \overline\gamma^\mu
\left\{ -\ii \overleftarrow \partial_\mu - e \, A_\mu \right\}
- \ii \, \Gamma_\mu \,  \overline\gamma^\mu - 
m_I \right] \, \gamma^0 \, \left[ \, \overline\psi(x) \, \right]^\plus \,.
\end{equation}
Now, we use the relations $\psi^\plus(x) \,  \gamma^0 = \overline\psi(x)$
and $ \gamma^0 \, \left[ \, \overline\psi(x) \, \right]^\plus = \psi(x)$,
and arrive at the form
\begin{equation}
\label{step1}
\calL^\plus = \overline\psi(x) \,
\left[  \overline\gamma^\mu
\left\{ -\ii \overleftarrow \partial_\mu - e \, A_\mu \right\}
- \ii \, \Gamma_\mu \,  \overline\gamma^\mu - 
m_I \right] \, \psi(x) \,.
\end{equation}
Because $\calL$ is a scalar, a transposition again does not
change the Lagrangian, and we have
\begin{equation}
\label{step2}
\left( \calL^\plus \right)^{\rm T} =  \psi^{\rm T}(x)  \,
\left[  \left( \overline\gamma^\mu \right)^{\rm T}
\left\{ -\ii \overrightarrow \partial_\mu - e \, A_\mu \right\}
- \ii \, \left( \overline\gamma^\mu \right)^{\rm T} \, 
\left( \Gamma_\mu \right)^{\rm T}  - m_I \right] \,
\left[ \overline\psi(x) \right]^{\rm T} \,.
\end{equation}
An insertion of the charge conjugation matrix $C = \ii \gamma^2 \, \gamma^0$
(with the flat-space $\gamma^2$ and $\gamma^0$)
leads to 
\begin{align}
\left( \calL^\plus \right)^{\rm T} =& \;  \psi^{\rm T}(x)  \, C^{-1} \,
\left[  C \, \left( \overline\gamma^\mu \right)^{\rm T} \, C^{-1} 
\left\{ -\ii \overrightarrow \partial_\mu - e \, A_\mu \right\}
\right.
\nonumber\\[0.1133ex]
& \; \left. 
- \ii \, C \, \left( \overline\gamma^\mu \right)^{\rm T} \, C^{-1} \, 
C \,  \Gamma_\mu^{\rm T}  \, C^{-1}  - m_I \right] \,
C \, \left[ \overline\psi(x) \right]^{\rm T} \,.
\end{align}
we use the identities 
$C \, \left( \overline\gamma^\mu \right)^{\rm T} \, C^{-1} = 
-\overline\gamma^\mu$, and $C \, \left( \Gamma_\mu \right)^{\rm T} \, C^{-1} = 
-\Gamma_\mu$. The latter of these can be shown as 
follows,
\begin{equation}
C \, \Gamma^{\rmT}_\mu \, C^{-1} = 
\frac{\ii}{4} \left\{ 
\frac{\ii}{2} \omega^{A B}_\mu \, 
C \, \left[ \gamma_B^{\rmT},\, \gamma_A^{\rmT} \right] \, C^{-1}
\right\}
= \frac{\ii}{4} \left\{ 
\frac{\ii}{2} \omega^{A B}_\mu \, 
\left[ -\gamma_B,\, -\gamma_A \right] 
\right\}
= -\Gamma_\mu \,.
\end{equation}
The result is the expression
\begin{equation}
\left( \calL^\plus \right)^{\rm T} =  \psi^{\rm T}(x)  \, C^{-1} \,
\left[  \left( -\overline\gamma^\mu \right) 
\left\{ -\ii \overrightarrow \partial_\mu - e \, A_\mu \right\}
- \ii \, \left( -\overline\gamma^\mu \right) \, 
\left( -\Gamma_\mu \right) - m_I \right] \,
C \, \left[ \overline\psi(x) \right]^{\rm T} \,.
\end{equation}
Now we express the result in terms of the 
charge-conjugate spinor $\psi^\calC(x)$ and its adjoint
$\overline{ \psi^\calC(x) }$
(further remarks on this point are presented in Appendix~\ref{appb}),
\begin{equation}
\label{psiC}
\psi^\calC(x) = C \, \left[ \overline\psi(x) \right]^{\rm T} \,,
\qquad
\overline{ \psi^\calC(x) }  = 
-\psi^{\rm T}(x)  \, C^{-1} \,,
\end{equation}
where we use the identity $C^{-1} = -C$ (see also Appendix~\ref{appa}). 
The Lagrangian becomes
\begin{align}
\label{Ltrafo}
\calL =& \; \left( \calL^\plus \right)^{\rm T} 
= -\overline{ \psi^\calC(x) } \,
\left[  \overline\gamma^\mu  
\left\{ \ii \overrightarrow \partial_\mu + e \, A_\mu \right\}
- \ii \, \overline\gamma^\mu \, \Gamma_\mu - m_I \right] \,
\psi^\calC(x) 
\nonumber\\[0.1133ex]
=& \; -\overline{ \psi^\calC(x) } \,
\left[  \overline\gamma^\mu  
\left\{ \ii (\partial_\mu - \Gamma_\mu) + 
e \, A_\mu \right\} - m_I \right] \,
\psi^\calC(x) \,.
\end{align}
The Lagrangian given in Eq.~\eqref{Ltrafo} differs from~\eqref{Lorig} 
only with respect to the sign of electric charge,
as is to be expected, and with respect to the replacement
of the Dirac spinor $\psi(x)$ by its charge conjugation
$\psi^\calC(x)$. The overall minus sign is physically irrelevant
as it does not influence the variational equations 
derived from the Lagrangian;
besides, it finds a natural explanation in terms of 
the reinterpretation principle,
if we interpret $\psi(x)$ as a Dirac wave function 
in first quantization.

Namely, there is a connection of the spatial integrals
of the mass term, proportional to 
\begin{equation}
\label{intJ}
J = \int \dd^3 r \, \overline\psi(x) \psi(x) =
\int \dd^3 r \, \overline\psi(t, \vec r) \, \psi(t, \vec r) 
= \int \dd^3 r \, \psi^\plus(t, \vec r) \gamma^0 \psi(t, \vec r) \,,
\end{equation}
and the charge conjugate,
\begin{equation}
\label{intJC}
J^\calC = \int \dd^3 r \, \overline\psi^\calC(x) \psi^\calC(x) =
\int \dd^3 r \, \left(\psi^\calC(t, \vec r) \right)^\plus
\gamma^0 \psi(t, \vec r) \,.
\end{equation}
Both of the above integrals connect to the 
energy eigenvalue of the 
Dirac equation in the limit of time-independent fields
(see Appendices~\ref{appa} and~\ref{appb}).
One can show that the energy eigenvalues of 
Dirac eigenstates $\psi$,
in the limit of weak 
potentials and states composed of small momentum components,
exactly correspond to the 
integrals $J$ and $ J^\calC$ (up to a factor $m_I$).
In turn, the dominant term in the Lagrangian 
in this limit is 
\begin{equation}
\calL \to - \overline{ \psi }(x) \, m_I \, \psi(x) 
= + \overline{ \psi^\calC(x) } \, m_I \, \psi^\calC(x) \,.
\end{equation}
Because the integral $\int \dd^3 r \, \calL$
equals $-J$ (or $+J^\calC$), the sign change becomes 
evident: it is due to the fact that the 
states $\psi^\calC$ describe antiparticle wave 
functions where the sign of the energy flips
in comparison to particles.
The matching of $m_I$ to the gravitational mass can 
be performed in a central, static field~\cite{JeNo2013pra,Je2019ijmpa},
and results in the identification $m_I = m_G$, 
where $m_G$ is the gravitational mass.
The gravitational covariant derivative
$\partial_\mu - \Gamma_\mu$ has retained its 
form in going from~\eqref{Lorig} to~\eqref{Ltrafo}, in agreement with the 
perfect particle-antiparticle symmetry of the 
gravitational interaction. Because the above demonstration
is general and holds for arbitrary (possibly dynamic) 
space-time background $\Gamma$, there is no room 
for a deviation of the gravitational interactions
of antiparticles (antimatter) to deviate from those of 
matter. This has been demonstrated here on the 
basis of Lagrangian methods, supplementing a recent
preliminary result~\cite{Je2019ijmpa}.

In order to fully clarify the origin of the minus sign
introduced upon charge conjugation,
one consults Chaps.~2 and 3 of Ref.~\cite{ItZu1980}
and Chap.~7 of Ref.~\cite{Kl2016}.
Namely, in second quantization, 
there is an additional minus sign incurred
upon the charge conjugation, which 
restores the original sign pattern of the 
Lagrangian.
According to Eq.~(2.107) and~(3.157) of Ref.~\cite{ItZu1980},
we can write the expansion of the 
free Dirac field operator as
\begin{equation}
\label{fieldop}
\hat\psi(x) = \sum_s \int \frac{\dd^3 p}{(2 \pi)^3} \,
\frac{m}{E} \,
\left[  a_s(\vec p) \, u_s(\vec p) \, \ee^{-\ii p \cdot x} 
+ \ee^{\ii p \cdot x} \,
v_s(\vec p) \, b^\plus_s(\vec p) 
\right] \,.
\end{equation}
The field operator is denoted by a hat 
in order to differentiate it from the 
Dirac wave function.
The four-momentum is $p^\mu = (E, \vec p)$,
where $E = \sqrt{\vec p^2 + m^2}$
is the free Dirac energy,
and $u_s(\vec p)$ and $v_s(\vec p)$ are the 
positive-energy and negative-energy spinors
with spin projection $s$ (onto the $z$ axis).
Furthermore, the particle annihilation operator
$a_s(\vec p)$ and the antiparticle 
creation operator $b^\plus_s(\vec p)$,
and their Hermitian adjoints, fulfill the 
commutation relations given in Eqs.~(3.161) of Ref.~\cite{ItZu1980},
\begin{subequations}
\begin{align}
\left\{ a_s(\vec p), a^\plus_{s'}(\vec p) \right\} =& \;
\frac{E}{m} \, (2 \pi)^3 \, \delta^{(3)}(\vec p - \vec p') \, \delta_{s s'} \,,
\\[0.1133ex]
\left\{ b_s(\vec p), b^\plus_{s'}(\vec p') \right\} =& \;
\frac{E}{m} \, (2 \pi)^3 \, \delta^{(3)}(\vec p - \vec p') \, \delta_{s s'} \,.
\end{align}
\end{subequations}
The spinors are normalized according to 
Eq.~(2.43a) of Ref.~\cite{ItZu1980},
i.e., they fulfill the relation
$ u^\plus_s(\vec p) \, u_s(\vec p) = 
v^\plus_s(\vec p) \, v_s(\vec p) = E/m$.
For the charge conjugation in the second-quantized theory,
it is essential that an additional 
minus sign is incurred in view of the 
anticommutativity of the field operators.
Namely, without considering the interchange of the 
field operators,
one would have, under charge conjugation,
$J^\mu(x) = \overline\psi(x) \gamma^\mu \psi(x) = 
\overline\psi^\calC(x) \gamma^\mu \psi^\calC(x) = 
J^{\calC \mu}(x)$, i.e., the 
current would not change under charge conjugation
which is intuitively inconsistent 
[see the remark following Eq.~(4.618) of Ref.~\cite{Kl2016}].
However, for the field operator current
(from here on, we denote field operators with a hat), we have
$\hat J^\mu(x) = \hat{\overline\psi}(x) \gamma^\mu \hat{\psi}(x) = 
- \hat{\overline\psi}{}^\calC(x) \gamma^\mu {\hat\psi}^\calC(x) = 
- \hat{J}^{\calC \mu}(x)$,
because one has incurred an additional 
minus sign due to the restoration of the 
field operators into their canonical order after charge
conjugation [see the remark following Eq.~(7.309) of Ref.~\cite{Kl2016}].

In our derivation above,
when one transforms to a second-quantized Dirac
field (but keeps classical background electromagnetic 
field and a classical non-quantized curved-space-time 
metric), one starts from Eq.~\eqref{step1} as an
equivalent, alternative formulation of Eq.~\eqref{ord}.
One observes that in going from Eq.~\eqref{step1} to~\eqref{step2},
one has actually changed the order of the 
field operators in relation to the 
Dirac spinors. Restoring the original order,
much in the spirit of Eq.~(7.309) of Ref.~\cite{Kl2016},
one incurs an additional minus sign which 
ensures that
\begin{align}
\label{theorem}
\hat{\calL} =& \; \hat{\overline \psi}(x)
\left[ \overline\gamma^\mu 
\left\{ \ii \left( \partial_\mu - \Gamma_\mu \right) - e \, A_\mu \right\} - 
m_I \right] \, \hat{\psi}(x) 
\nonumber\\[0.1133ex]
=& \; \hat{\overline \psi}{}^\calC(x) \,
\left[  \overline\gamma^\mu
\left\{ \ii (\partial_\mu - \Gamma_\mu) +
e \, A_\mu \right\} - m_I \right] \,
\hat{\psi}^\calC(x) \,,
\end{align}
exhibiting the effect of charge conjugation 
in the second-quantized theory---and restoring the 
overall sign of the Lagrangian.
The theorem~\eqref{theorem} shows that particles and 
antiparticles behave exactly the same in 
gravitational fields, but it does not imply,
{\em a priori}, that $m_I = m_G$. 
The matching of the inertial mass $m_I$ and the 
gravitational mass $m_G$ most easily proceeds in a 
central, static field (Schwarzschild metric),
as demonstrated in Sec.~3 of Ref.~\cite{Je2019ijmpa}.

One should, at this stage, remember that 
experimental evidence, to the extent possible,
supports the above derived symmetry relation.
The only direct experimental result on antimatter and
gravity comes, somewhat surprisingly,
from the Supernova 1987A. 
Originating from the Large Magellanic Cloud, the originating
neutrinos and antineutrinos
eventually were detected on Earth. 
In view of their travel time of about 
160,000 years, they were bent from a ``straight line''
by the gravity from our own galaxy. 
The gravitational bending changed the time needed
to reach Earth by about 5 months.
Yet, both neutrinos and antineutrinos
reached Earth within the same 12 second interval,
shows that neutrinos and antineutrinos fall similarly,
to a precision of about 1 part in a million~\cite{Lo1988,LS1988}.
In view of the exceedingly small rest mass
of neutrinos, the influence of the 
mass term (even a conceivable tachyonic mass term)
on the trajectory is negligible~\cite{NoJe2015tach}.
Yet, it is reassuring that 
experimental evidence, at this time,
is consistent with Eq.~\eqref{theorem}.

%
%
\section{Other Interpretations of Antimatter Gravity}
\label{sec4}

In view of the symmetry relations derived 
in this article for the gravitationally and electromagnetically
coupled Dirac equation, it is 
certain justified to ask about an adequate interpretation 
of antimatter gravity experiments.
We have shown that canonical gravity cannot 
account for any deviations of gravitational interactions 
of matter versus antimatter.
How could tests of antimatter ``gravity'' 
be interpreted otherwise?
The answer to that question involves clarification
of the question which ``new'' interactions could
possibly mimic gravity. 
The criteria are as follows:
{\em (i)} The ``new'' interaction would 
need to violate CPT symmetry.
{\em (ii)} The ``new'' interaction 
would have to be a long-range
interaction, mediated by a massless 
virtual particle. 

One example of such an interaction would be induced if 
hydrogen atoms were to acquire, in addition to the 
electric charges of the constituents (electrons and protons),
an additional ``charge'' $\eta \, e$,
where $e$ is the elementary charge, while antihydrogen
atoms would acquire a charge $-\eta \, e$,
where $\eta$ is a small parameter.
One could conjecture the existence of a small,
CPT-violating ``charge'' $\eta e/2$ for electrons,
protons, and neutrons, while positrons and antiprotons,
and antineutrons, would carry a ``charge'' $-\eta e/2$.
We will refer to this concept as the ``$\eta$ force''
in the following. The difference in the gravitational force
(acceleration due to the Earth's field)
felt by a hydrogen versus an antihydrogen atom is 
\begin{equation}
F^\eta_{\overline{\rm H}} - F^\eta_{{\rm H}} =
2 \, \eta \, \left[ \frac{\eta}{2} (N_p + N_n + N_e) \right] \,
\frac{e^2}{4 \pi \epsilon \, R_\oplus^2} \,.
\end{equation}
Here, $R_\oplus$ is the radius of the Earth,
while $N_p$, $N_n$ and $N_e$ are the numbers of 
protons, neutrons and electrons in the Earth.
The gravitational force on a falling antihydrogen
atom is 
\begin{equation}
F^G_{\overline{\rm H}} =
G \, \frac{m_p \, M_\oplus}{R_\oplus^2}  \,.
\end{equation}
Let us assume that
an experiment establishes that 
$| F^\eta_{\overline{\rm H}} - F^\eta_{{\rm H}} | <
\chi \, F^G_{\overline{\rm H}}$, where $\chi$ 
is a measure of the deviation of the acceleration
due to gravity$+$``$\eta$''-force
for antihydrogen versus hydrogen.
A quick calculation shows that this translates into 
a bound
\begin{equation}
\eta < 7.3 \times 10^{-19}\, \sqrt{\chi} \,.
\end{equation}
Antimatter gravity tests thus limit the available 
parameter space for $\eta$, and could be interpreted in
terms of corresponding limits on the maximum allowed 
value of $\eta$.

%
%
\section{Conclusions}
\label{sec5}

In the current paper, we have analyzed the 
particle-antiparticle symmetry of the 
gravitationally (and electromagnetically) 
coupled Dirac equation and come to the conclusion
that a symmetry exists, for the second-quantized
formulation, which precludes the existence
particle-antiparticle symmetry breaking terms
on the level of Dirac theory.
In a nutshell, one might say the following:
Just as much as the electromagnetically coupled Dirac equation
predicts that antiparticles have the opposite charge as compared
to particles (but otherwise behave exactly the same under
electromagnetic interactions),
the gravitationally coupled Dirac equation predicts that
particles and antiparticles follow exactly the same dynamics in curved space-time,
i.e., with respect to gravitational
fields (in particular, they have the same gravitational mass, and there is no
sign change in the gravitational coupling).
In the derivation of our theorem~\eqref{theorem}, 
we use the second-quantized Dirac formalism,
in the Lagrangian formulation.
Our general result for the Dirac adjoint,
communicated in Sec.~\ref{sec2},
paves the way for the Lagrangian of the 
gravitationally coupled field, and its explicit form
is an essential ingredient of our considerations.

Why is this interesting? Well, first, because the 
transformation of the gravitational force
under the particle-to-antiparticle transformation 
has been discussed controversially in the 
literature~\cite{Sa1999,Vi2011,Ca2011comment,Vi2011reply}.
In Ref.~\cite{HoEtAl1991}, it was pointed out that 
the role of the CPT transformation in gravity needs
to be considered with care: It relates the 
fall of an apple on Earth to the fall of an 
anti-apple on anti-Earth, but not on Earth.
The Dirac equation, colloquially speaking,
applies to both apples as well as anti-apples on Earth, i.e., 
to particles and antiparticles in the same space-time metric.
Second, our results have important consequences because
one might have otherwise speculated about the 
existence of tiny violations of the 
particle-antiparticle symmetry, even on the level
of the gravitationally coupled Dirac theory.
For example, in Ref.~\cite{Ob2001}, it was claimed 
that the Dirac Hamiltonian for a particle
in a central gravitational field, after a Foldy--Wouthuysen
transformation which disentangles the particle
from the antiparticle degrees of freedom,
contains the term
[see the last term on the first line of the 
right-hand side of Eq.~(31)]
\begin{equation}
\label{Hsim}
H \sim - \frac{\hbar}{2 c} \vec \Sigma \cdot \vec g \,,
\qquad
\vec \Sigma = 
\left( \begin{array}{cc} \vec\sigma & 0 \\
0 & \vec\sigma \end{array} \right) \,.
\end{equation}
We here restore the factors $\hbar$ and $c$ in order
to facilitate the comparison to Ref.~\cite{Ob2001}.
The term proportional to $\vec \Sigma \cdot \vec g$,
where $\vec g$ is the acceleration due to gravity,
would break parity, because $\vec \Sigma$ 
transforms as a pseudovector, while $\vec g$ 
transforms as a vector under parity.
This aspect has given rise to discussion,
based on the observation that an initially parity-even
Hamiltonian (in a central field) should not give
rise to parity-breaking terms after
a disentangling of the effective Hamiltonians
for particles and antiparticles~\cite{Ni2002,Ob2002}.

We should note that Ref.~\cite{Ob2001}
was not the only place in the literature where
the authors speculated about the existence of 
P, and CP--violating terms obtained after
the identification of low-energy operators 
obtained from Dirac Hamiltonians
in gravitational fields.
E.g., in Eq.~(46) of Ref.~\cite{DoHo1986},
spurious parity-violating, and CP-violating terms were obtained after a
Foldy--Wouthuysen transformation;
these terms would of course also violate particle-antiparticle
symmetry.

In the context of the current discussion,
the existence of terms proportional to 
$\vec \Sigma \cdot \vec g$, as given in Eq.~\eqref{Hsim},
would also violate particle-antiparticle symmetry:
This is because it lacks the universal prefactor 
$\beta = \gamma^0$, where
\begin{equation}
\beta = \left( \begin{array}{cc} 
\mathbbm{1}_{2 \times 2} & 0 \\
0 & -\mathbbm{1}_{2 \times 2} \end{array} 
\right) \,.
\end{equation}
In fact, in the complete result (up to fourth order in the 
momenta) for the effective
particle-antiparticle Hamiltonian in a central 
field, given in Eq.~(21) of Ref.~\cite{JeNo2013pra},
all terms have a common prefactor $\beta$.
The common prefactor $\beta$ implies that,
after the application of the reinterpretation
principle for antiparticles, 
the effective Hamiltonians for particles
and antiparticles in a central gravitational
field (but without electromagnetic coupling)
are exactly the same, and ensures the particle-antiparticle 
symmetry.

The absence of such parity-violating 
(and particle-antiparticle symmetry breaking)
terms has meanwhile been confirmed
in remarks following Eq.~(15) of Ref.~\cite{SiTe2005},
in the text following Eq.~(35) of Ref.~\cite{Si2016},
and also, in clarifying remarks given in the text following
Eq.~(7.33) of Ref.~\cite{ObSiTe2017}.
Further clarifying analyses can be found in
Ref.~\cite{JeNo2014jpa} and in Ref.~\cite{GoNe2014}.
Related calculations have
recently been considered in
other contexts~\cite{ObSiTe2014,ObSiTe2016,ObSiTe2017}.
The question of whether such parity-
and particle-antiparticle symmetry violating terms could exist
in higher orders in the momentum expansion
has been answered negatively in Ref.~\cite{Je2013},
but only for a static central gravitational field,
and in Ref.~\cite{NoJe2016}, still
negatively, for combined {\em static}, central
gravitational and electrostatic fields.

However, the question regarding the absence of 
particle-antiparticle symmetry breaking 
terms for general, dynamic space-time backgrounds 
has not been answered conclusively in the literature
up to this point, to the best of our knowledge.
This has been the task of the current paper.
In particular, our results imply a {\em no-go theorem regarding 
the possible emergence of particle-antiparticle-symmetry
breaking gravitational, and combined electromagnetic-gravitational
terms in general static and dynamic curved-space-time backgrounds.}
Any speculation~\cite{DoHo1986,Ob2001} about the 
re-emergence of such terms in a dynamic space-time background
can thus be laid to rest.
Concomitantly, we demonstrate that there are 
no ``overlap'' or ``interference'' terms 
generated in the particle-antiparticle transformation,
between the gauge groups,
namely, the $SO(1,3)$ gauge group of the local
Lorentz transformations, and the $U(1)$ gauge 
group of the electromagnetic theory.
This result implies both progress 
and, unfortunately, some disappointment, 
because the emergence of such terms would have 
been fascinating and would have opened up, quite possibly, interesting
experimental opportunities.
In our opinion, antimatter gravity experiments should be interpreted
in terms of limits on CPT-violating parameters, such as the 
$\eta$ parameter introduced in Sec.~\ref{sec4}.
This may be somewhat less exciting than a ``probe of the equivalence principle
for antiparticles'' but still, of utmost value for the 
scientific community.

\section*{Acknowledgments}

The author acknowledges support from the
National Science Foundation (Grant PHY--1710856)
as well as insightful conversations with J.~H.~Noble.

\appendix

%
%
\section{Sign Change of 
\texorpdfstring{$\maybebm{\overline \psi \, \psi}$}{psi-bar psi} 
under Charge Conjugation}
\label{appa}

With the charge conjugation matrix $C = \ii \gamma^2 \gamma^0$
(superscripts denote Cartesian indices), 
and the Dirac adjoint $\overline \psi = \psi^\plus \, \gamma^0$, we have
\begin{equation}
\psi^\calC = C \, \overline \psi^\rmT = 
\ii \gamma^2 \, \gamma^0 \, \gamma^0 \, \psi^* = \ii \gamma^2 \, \psi^* \,.
\end{equation}
We recall that the $\gamma^2$ (contravariant index, no square) 
matrix in the Dirac representation matrix is 
\begin{equation}
\gamma^2 = 
\left( \begin{array}{cc} 0 & \sigma^2 \\
-\sigma^2 & 0 \end{array} \right) \,,
\quad
\sigma^2 = \left( \begin{array}{cc} 0 & -\ii \\
\ii & 0 \end{array} \right) \,,
\quad
\left( \sigma^2 \right)^\plus = \sigma^2 \,,
\end{equation}
which implies that $\left( \gamma^2 \right)^\plus = -\gamma^2$.
The Dirac adjoint of the charge conjugate is
\begin{equation}
\overline \psi^\calC = 
\left( \psi^\calC \right)^\plus \, \gamma^0 = 
\psi^\rmT (-\ii) \left( \gamma^2 \right)^\plus \, \gamma^0 
= \psi^\rmT (-\ii) \, (-\gamma^2) \, \gamma^0 = 
\psi^\rmT \, \ii \, \gamma^2 \, \gamma^0 \,.
\end{equation}
This leads to a verification of the sign flip of
the mass terms in the gravitationally coupled
Lagrangian for antimatter, given in Eq.~\eqref{Ltrafo}
[see also Eqs.~\eqref{intJ} and~\eqref{intJC}],
\begin{equation}
\overline \psi^\calC \, \psi^\calC = 
( \psi^\rmT \ii \gamma^2 ) \, \gamma^0 \, ( \ii \, \gamma^2 \, \psi^* )
= -(\ii)^2 \psi^\rmT \, (\gamma^2)^2 \, \gamma^0 \psi^* 
= - \psi^\rmT \, \gamma^0 \, \psi^* 
= - \overline\psi \, \psi \,.
\end{equation}

Two useful identities {\em (i)} $\gamma^0 \, C^\plus \,  \gamma^0  = C$ and
{\em (ii)} $C^{-1} = -C$ have been used in Sec.~\ref{sec3}.
These will be derived in the following.
The explicit form of the $\gamma^2$ matrix in the Dirac
representation implies that
$\left( \gamma^2 \right)^\plus = -\gamma^2$.
Based on this relation, we can easily show that
\begin{equation}
C^\plus = 
\left( \ii \, \gamma^2 \, \gamma^0 \right)^\plus =
-\ii \, \gamma^0 \, \left( \gamma^2 \right)^\plus 
= \ii \, \gamma^0 \, \gamma^2 = 
-\ii \, \gamma^2 \, \gamma^0 = -C \,.
\end{equation}
The first identity $\gamma^0 \, C^\plus \,  \gamma^0  = C$ can
now be shown as follows,
\begin{equation}
\gamma^0 \, C^\plus \,  \gamma^0 =
\gamma^0 \, \left[ -\ii \, \gamma^2 \, \gamma^0 \right] \,  \gamma^0 =
-\ii \, \gamma^0 \, \gamma^2 = \ii \, \gamma^2 \, \gamma^0 = C \,.
\end{equation}
Furthermore, one has
\begin{equation}
C \, C^\plus = 
C \, (-C) =  
 \ii \, \gamma^2 \, \gamma^0 \;
 \ii \, \gamma^0 \, \gamma^2 
= - \left( \gamma^2 \right)^2 
= - \left( -\mathbbm{1}_{4 \times 4} \right)
= \mathbbm{1}_{4 \times 4} \,,
\end{equation}
so that 
\begin{equation}
C^{-1} =  C^\plus = - C \,,
\end{equation}
which proves, in particular, that $C^{-1} = -C$.

%
%
\section{General Considerations}
\label{appb}

A few illustrative remarks are in order.
These concern the following questions:
{\em (i)} To which extent do gravitational and 
electrostatic interactions differ 
for relativistic particles?
This question is relevant because, in the nonrelativistic 
limit, in a central field, both interactions are 
described by potentials of the same 
functional form (``$1/R$ potentials''). 
{\em (ii)} Also, we should clarify why the 
integrals~\eqref{intJ} and~\eqref{intJC}
represent the dominant terms in the 
evaluation of the Dirac particle energies,
in the nonrelativistic limit.

After some rather deliberate and extensive 
considerations, one can show~\cite{Je2018geonium}
that, up to corrections which combine 
momentum operators and potentials, the general Hamiltonian for a 
Dirac particle in a combined 
electric and gravitational field is
\begin{equation} 
H_D = \vec\alpha \cdot \vec p + \beta \{ m (1 + \phi_G) \} + e \phi_C \,,
\end{equation} 
where $\phi_G$ is the gravitational, and $\phi_C$ is the 
electrostatic potential.
Also, $\vec\alpha$ is the vector of Dirac $\alpha$ matrices,
$\vec p$ is the momentum operator, 
and $\beta = \gamma^0$ is the Dirac $\beta$ matrix.
After a Foldy--Wouthuysen transformation~\cite{JeNo2014jpa},
one sees that the gravitational interaction respects the 
particle-antiparticle symmetry, while 
the Coulomb potential does not, commensurate with the 
opposite sign of the charge for antiparticles.
Question {\em (i)} as posed above can thus be answered 
with reference to the fact that, in leading approximation,
the gravitational potential enters the Dirac equation 
as a scalar potential, modifying the mass term,
while the electrostatic potential 
can be added to the free Dirac Hamiltonian $vec\alpha \cdot \vec p + \beta m$
by covariant coupling~\cite{ItZu1980}.

The second question posed above is now easy to answer:
Namely, in the nonrelativistic limit, one has 
\begin{equation} 
\vec\alpha \cdot \vec p \to 0 \,,
\end{equation} 
and furthermore, the gravitational and electrostatic potentials
can be assumed to be weak against the mass term,
at least for non-extreme Coulomb fields~\cite{MoPlSo1998}.
Under these assumptions, one has $H_D \to \beta m$, and 
the matrix element $\langle \psi | H_D | \psi\rangle$ assumes the form 
$\int \dd^3 r \, \psi^\plus(\vec r) \, \gamma^0 \, m \psi(\vec r)$
[see Eq.~\eqref{intJ}].

\end{document}